\documentstyle[12pt]{article}
\def\theequation{\arabic{section}.\arabic{equation}}
\def\appendix{
\vskip 1cm
\par
\setcounter{equation}{0}
\def\theequation{A1.\arabic{equation}}
}
\begin{document}
\begin{titlepage}
\begin{center}
\hspace*{10cm} Preprint IFUNAM\\
\hspace*{10cm}  FT-93-35\\
\hspace*{10cm}  November 1993\\
\vspace*{15mm}
{\Large {\bf Gluonium as Bound State of  Massive Gluons\\
\vskip1mm Described by the Joos-Weinberg
Wave Functions.}}\\
\vspace{0.3cm}
{\tt  VALERI  V.  DVOEGLAZOV $^{*,\,\dagger}$}\\
\vskip 1mm
{\it  Departamento de F\'{\i}sica Te\'{o}rica, \,Instituto de F\'{\i}sica,\\
Universidad Nacional Aut\'{o}noma de M\'{e}xico, \\
Apartado Postal 20-364, 01000 D.F. , MEXICO}\\
\vskip2mm
{\tt and}\\
\vskip2mm
{\tt SERGEI V. KHUDYAKOV $^{*,\,\ddagger}$}\\
 \vskip 1mm
{\it  Laboratory of Theoretical Physics\\
Joint Institute for Nuclear Research\\
Head Post Office, P. O. Box 79,
Moscow 101000  RUSSIA}\\
\end{center}
\vspace*{3mm}
\small{
\noindent
{\tt ABSTRACT.} On the basis of the Kadyshevsky equal-time (quasipotential)
approach, a set of partial-wave equations is derived for the wave function
of a gluonium, a bound state of two gluons. The field operators
of  constituent gluons are considered as six component quantities
according to the Joos-Weinberg  $2(2S+1)$- component approach.
The quasiclassical quantization condition
 for relativistic two-particle states and the above
set  can be  used  for calculations of   gluonium energy levels.\\

\noindent
{\tt RESUMEN.} Partiendo del m\'{e}todo de tiempos iguales (cuasi-potencial) de
Kadyshevsky se deduce un conjunto de ecuaciones de ondas parciales para
la funci\'{o}n de onda del gluonio, el estado ligado de dos gluones. Los
operadores
de campo de los gluones constituyentes se consideran cantidades de seis
componentes de acuerdo con el modelo de Joos-Weinberg de 2 (2S + 1)
componentes. La condici\'{o}n de cuantizaci\'{o}n cuasi-cl\'{a}sica para  los
estados
relativistas de dos part\'{\i}culas y el conjunto de ecuaciones mencionado
pueden emplearse para calcular los niveles de energ\'{\i}a del gluonio.
}

\vspace*{5mm}
\vspace*{5mm}
(glueball), Joos-Weinberg approach

\vspace*{-5mm}
\noindent

\noindent
\footnotesize{
$^{*}$ On leave of absence from {\it Dept.Theor.} \& {\it Nucl. Phys.,
   Saratov State University and Sci.} \& {\it Tech. Center for  Control
and Use of Physical Fields and Radiations, Astrakhanskaya str. , 83,\,\,
   Saratov 410071 RUSSIA}\\
$^{\dagger}$ Email: valeri@ifunam.ifisicacu.unam.mx,\,
dvoeglazov@main1.jinr.dubna.su\\
$^{\ddagger}$ Email: khud@theor.jinrc.dubna.su,\, vapr@scnit.saratov.su }
\end{titlepage}

\setcounter{equation}{0}
\section{Introduction}

\hspace*{8mm}The existence of a gluonium, which is a color singlet of the bound
state of two or more gluons, is predicted by all the models of quantum
chromodynamics (QCD), the lattice models~\cite{a1},
the sum rules~\cite{a2}, the bag model~\cite{a3} and the effective Lagrangian
approach~\cite{a3a}.  Experimental searches of these states are in progress,
see, e. g., for the reviews in ref.~\cite{a4}. It follows from the analysis of
obtained results that the most likely candidates
for glueballs  are the following meson resonances~\cite{a51c}:\\

Table. {\it The eventual candidates for glueball.}

\vspace*{5mm}

\begin{tabular}{||l|c|c|c|l||}
\hline
\hline
     &References&$J^{PC}$&$n\,^{2S+1}L_{J}$&Reactions of observations\\
\hline
$\sigma(750)$&\cite{e1}&$0^{++}$&$1\,^{1}S_{0}$&$\pi\pm
N_{pol}\rightarrow\sigma
N$\\
\hline
$\imath(1460)$&\cite{e8}&$0^{-+}$&$1\,^{3}P_{0}$&$J/\psi\rightarrow\gamma X$\\
\hline
$G(1590)$&\cite{e3}&$0^{++}$&$2\,^{1}S_{0}$&$\pi^{-}p\rightarrow\eta\eta n$\\
\hline
$\vartheta(1720)$&\cite{e2,e7}&$2^{++}$&$2\,^{5}S_{2}$&$J/\psi\rightarrow\gamma
X$\\
\hline
$g(2050)$&\cite{e71}-\cite{e6}&$2^{++}$&$3\,^{5}S_{2}$
&$\pi^{-}p\rightarrow\phi\phi n, \quad J/\psi\rightarrow\gamma X$\\
\hline
$\xi(2220)$&\cite{e71}&$2^{++}$&$2\,^{3}D_{2}\quad or\quad
2\,^{5}D_{2}$&$J/\psi\rightarrow\gamma X$\\
\hline
$g'(2300)$&\cite{e71}-\cite{e6}&$2^{++}$&$3\,^{5}D_{2}$
&$\pi^{-}p\rightarrow\phi\phi n,\quad J/\psi\rightarrow\gamma X$\\
\hline
$g''(2350)$&\cite{e71}-\cite{e6}&$2^{++}$&$3\,^{1}D_{2}$
&$\pi^{-}p\rightarrow\phi\phi n,\quad J/\psi\rightarrow\gamma X$\\
\hline
\hline
\end{tabular}

\vspace*{10mm}

In connection with that it is very important to describe the gluonium spectra
theoretically.  Attempts have been made earlier to consider gluonium in the
framework of the potential model with massive structure gluons ~\cite{a51c,a51}
analogous to  the non-relativistic description of the quark-antiquark system,
ref.~\cite{a9}.

At present,  the relativistic single-time approach~\cite{a10}-\cite{a12} is
used widely for  the description of  two-particle systems like quarkonium. The
necessity of  allowance for
 relativistic effects is caused by the fact that in many cases the contribution
of the relativistic corrections is of the same order as the contribution of the
non-relativistic Hamiltonian. In the present work,
the quasipotential equal-time approach is employed
for the description of two-gluon bound states \footnote{The two-gluon bound
system has
 recently been described on the basis of the Bethe-Salpeter equation,
ref.~\cite{t6}.}
consisting of the structural gluons which are described by the six-component
Weinberg's wave functions~\cite{Joos}-\cite{Ahlu}. The results of the
papers~\cite{a14,a25},
 devoted to the covariant three-dimensional description of the composite system
formed by two particles with the $S=1$ spin, are used. The remarkable feature
of our formalism is the locality of the corresponding quasipotential in the
Lobachevsky momentum space. This is achieved by the separation of the
kinematical Wigner rotations and ''resetting''  all spin indices to the one
momentum,  for details see refs.~\cite{a11,Chesh,a20}.
Moreover, the quasipotential for interaction of two vector particles is the
same as the quasipotential for interaction of two spinor particles with
corresponding substitutions accounting for the spin difference and
the normalization.

The  quasipotential equation  with the one-boson exchange potential, obtained
in  Section II,  is reduced to the finite-difference partial-wave equations in
the relativistic configurational representation (RCR) in  Section III.  As was
shown earlier~\cite{WKB},  it is possible to develop the relativistic analog of
the WKB methods in the RCR.
This method has been successfully used to find  the quarkonium mass
spectrum~\cite{a12}. The use of the quasiclassical
quantization condition for relativistic two-particle states~\cite[a]{a12}, see
Section IV,
also allows  calculation the gluonium energy levels.

\setcounter{equation}{0}
\section{Spin structure of the relativistic potential for two-gluon interaction
 in the momentum representation}

\hspace*{8mm}At present,  the gluon could be described as a massive particle
with   dynamical mass appearing due to the existence
of  color charge and  self-interaction. This fact permits one to eliminate some
contradictions in the results of calculations of the proton formfactor and the
effective coupling constant $\alpha_S(q^2)$
on the basis of QCD (see in this connection ref.~\cite{a15}).

Therefore,  we begin by considering the quasipotential for two-gluon
interaction in the momentum representation as
that of gluonium consisting of  the structure massive gluons with the
intermediate interaction of the gauge massless gluon. The corresponding diagram
describing this process is drawn at  {\it Fig. 1 }. The Feynman matrix element
$<~p_1, p_2; \sigma_1, \sigma_2\mid
\hat{T}^{(2)}\mid k_1, k_2; \nu_1, \nu_2~>$ corresponding to this diagram is
considered  to be the quasipotential, $\hat{V}^{(2)}=\hat{T}^{(2)}$.

\vspace*{3mm}
\let\picnaturalsize=N
\def\picsize{3.2in}
\def\picfilename{gl1.eps}
\ifx\nopictures Y\else{\ifx\epsfloaded Y\else\input epsf \fi
\let\epsfloaded=Y
\centerline{\ifx\picnaturalsize N\epsfxsize \picsize\fi
\epsfbox{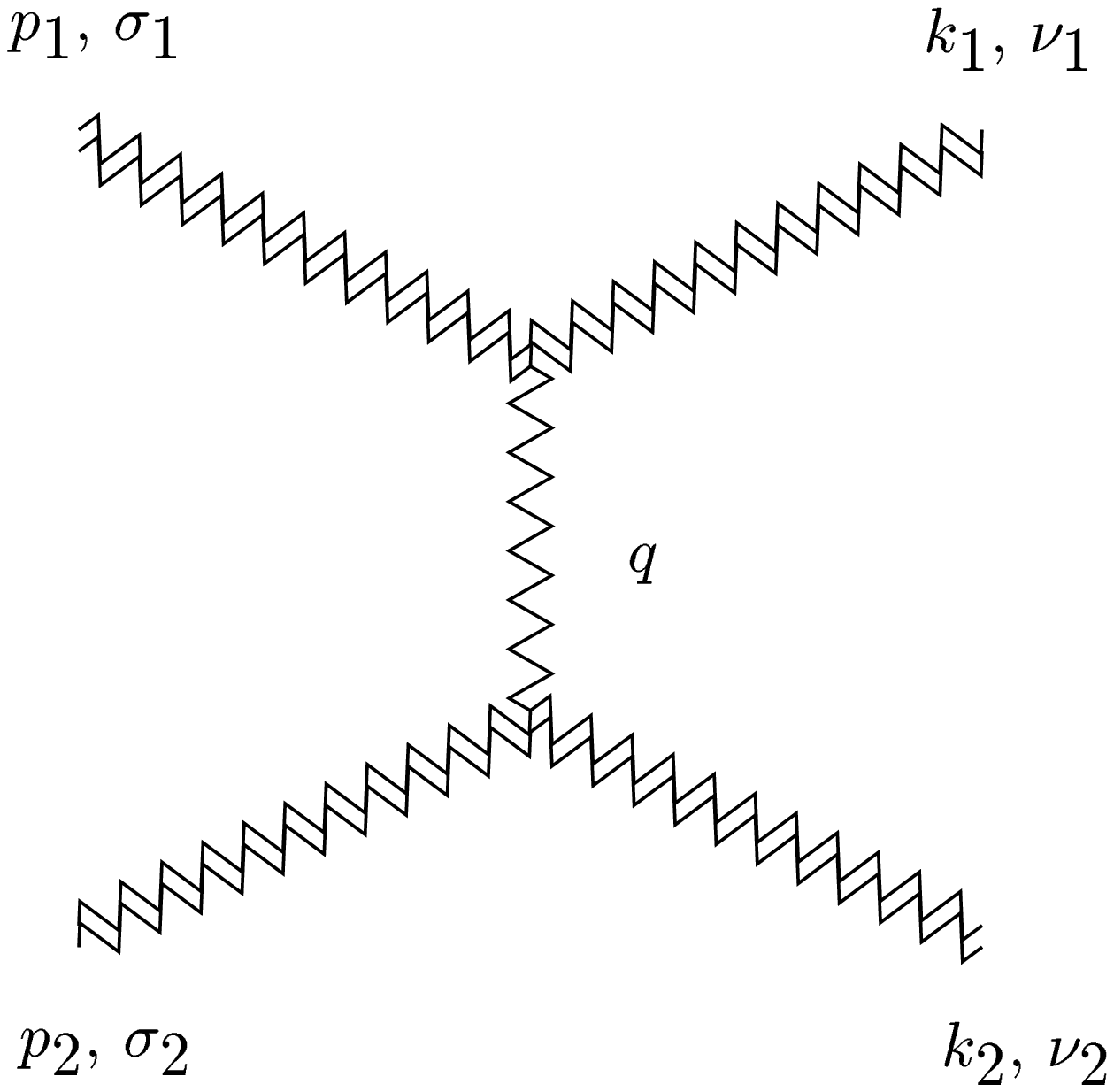}}}\fi

To find  the form of the quasipotential it is necessary to know the
Feynman rules for the vertices of interaction of the structure gluons and the
massless gauge gluon.

In  ref.~\cite{Joos} the attractive $2(2S+1)$- formalism for the description of
 particles of spin  $S=1$ has been proposed. As opposed to the Proca functions,
which transform according to the $({1\over 2}, {1\over 2})$ representation of
the Lorentz group in the case of $S=1$, the spinor functions are constructed
via the representation $(S,0)\oplus (0,S)$ in the Joos-Weinberg formalism. This
way of description of higher spin particles is on an equal footing to the Dirac
description of spinor particles whose wave functions transform according to the
$({1\over 2},0)\oplus (0, {1\over 2})$ representation. The $2(2S+1)$- component
analogues of the Dirac functions in the momentum space are\footnote{These
functions obeys the orthonormalization
equations, $U^+(\vec p)\gamma_{44} U(\vec p)=1 $ and   analogous equation
exists for  $V(\vec p)$, the functions of  negative-energy states.}
\begin{equation}
U(\vec p)={1\over\sqrt{2}}\left (\matrix{
D^S \left (\alpha(\vec p)\right )\xi_\sigma\cr
D^S \left (\alpha^{-1\,+}(\vec p)\right )\xi_\sigma\cr
}\right ),
\end{equation}
for the positive-energy states; and
\begin{equation}
V(\vec p)={1\over \sqrt{2}}\left (\matrix{
D^S \left (\alpha(\vec p)C^{-1}\right )\xi^*_\sigma\cr
D^S \left ( \alpha^{-1\,+}(\vec p)C^{-1}\right )(-1)^{2S}\xi^*_\sigma\cr
}\right ),
\end{equation}
for the negative-energy states, with the following notations:
\begin{equation}
\alpha(\vec p)=\frac{p_0+M+(\vec\sigma\vec p)}{\sqrt{2M(p_0+M)}},\quad
C=-i\sigma_2;
\end{equation}
and $D^S (A)\equiv D^{(S,0)} (A)$ is the Lorentz group representation by
matrices  with  $(2S+1)$ rows and columns\footnote{The technique of
construction of $D^S(A)$  could be found in~\cite[k]{Joos}.}. In the case of
$S=1$, one has
\begin{equation}
D^{(1,0)}\left (\alpha(\vec p)\right ) = 1+\frac{(\vec S\vec p)}{M}+\frac{(\vec
S\vec p)^2}{M(p_0+M)}.
\end{equation}

In spite of some antiquity of this formalism, in our opinion, it does not
deserve to be retired. Recently, some attention has been paid to this
formalism~\cite{Good}-\cite{a25}.

In the articles~\cite[b,h-j]{Joos}  the Feynman diagram technique is discussed
for the vector particles in the above-mentioned six-component formalism for
quantum electrodynamics (QED). The following Lagrangian:
\begin{eqnarray}
\lefteqn{{\cal L}^{QED}=\bar{\Psi}(x)\Gamma_{\mu\nu}{\bar \nabla}_{\mu}
\vec{\nabla}_{\nu}\Psi(x)-M^2\bar{\Psi}(x)\Psi(x)-{1 \over
4}F_{\mu\nu}F_{\mu\nu}+}\nonumber\\
&+&\frac{e\lambda}{12}F_{\mu\nu}\bar{\Psi}(x)\gamma_{5,\mu\nu}\Psi(x)+\frac{e
\kappa}{12 M^2}\partial_{\alpha}F_{\mu\nu}\bar{\Psi}(x)
\gamma_{6,\mu\nu,\alpha\beta}\nabla_{\beta}\Psi(x)
\end{eqnarray}
has been  used there.
In the above formula we have $\nabla_{\mu}=-i\partial_{\mu}-eA_{\mu}$;
$F_{\mu\nu}=\partial_{\mu}A_{\nu}-\partial_{\nu}A_{\mu}$ is the
 electromagnetic field tensor; $A_{\mu}$ is the 4- vector of the
electromagnetic field; $\bar{\Psi}, \Psi$  are the six-component wave functions
(WF) of the massive vector particle. The following expression has been obtained
for the interaction vertex of a vector particle with a
photon~\cite[j]{Joos},\cite{a25}:
\begin{equation}
-e\Gamma_{\alpha\beta}(p+k)_{\beta}-{ie\lambda \over
6}\gamma_{5,\alpha\beta}q_{\beta}+{e\kappa \over
6M^2}\gamma_{6,\alpha\beta,\mu\nu}q_{\beta}q_{\mu}(p+k)_{\nu},
\label{22}
\end{equation}
where $\Gamma_{\alpha\beta}=\gamma_{\alpha\beta}+\delta_{\alpha\beta}$;
$\gamma_{\alpha\beta}$; $\gamma_{5, \alpha\beta}$; $\gamma_{6,\alpha\beta,
\mu\nu}$ are $6\otimes 6$- matrices which have been  considered  in
ref.~\cite[b,g]{Joos}

\begin{center}
$\gamma_{ij}=\pmatrix{
0 & \delta_{ij}-S_i S_j- S_j S_i \cr
\delta_{ij}-S_i S_j- S_j S_i & 0 \cr
}$\\
\vspace*{4mm}
$\gamma_{i4}=\gamma_{4i}=\pmatrix{
0 & iS_i \cr
-iS_i & 0 \cr
}$,
$\gamma_{44}=\pmatrix{
0 & 1 \cr
1 & 0\cr
}$,\\
\end{center}
\vspace*{5mm}
(here, $S_i$ are the  spin matrices for a vector particle),
\begin{eqnarray}
\gamma_{5,\alpha\beta}&=&i [\gamma_{\alpha\mu}, \gamma_{\beta\mu}],\\
\gamma_{6,\alpha\beta,\mu\nu}&=&
[\gamma_{\alpha\mu},\gamma_{\beta\nu}]_{+}
+2\delta_{\alpha\mu}\delta_{\beta\nu}-[\gamma_{\beta\mu},
\gamma_{\alpha\nu}]_{+}-2\delta_{\beta\mu}\delta_{\alpha\nu},
\end{eqnarray}
$e$ is  electron charge, $\lambda$ and $\kappa$ are the quantities which
correspond to the magnetic dipole moment and the electric quadrupole moment,
respectively; $M$ is the vector particle mass.

In the case of interaction of two structure gluons we also use the $2(2S+1)$-
component formalism. Since the gluon $2(2S+1)$- dimensional WF can not be
directly introduced into the Lagrangian by means of the standard procedure of
lengthening the derivative (covariantization),  we add the terms defining the
structure gluons manually:
\begin{equation}
{\cal L}^{QCD}_{aux}=\bar{g}{\bar\nabla}_{\mu}\gamma^{\mu\nu}\vec \nabla_{\nu}
g-M^2_g \bar g g
\end{equation}
into the commonly used QCD Lagrangian:
\begin{equation}
{\cal L}^{QCD}=i\bar q \gamma^{\mu}\nabla_{\mu}q - m_{q}\bar q q -{1 \over
4}G^a_{\mu\nu}G^{a,\mu\nu}.
\end{equation}
Where
\begin{equation}
\nabla_{\mu}g(x)=\partial_{\mu}g-ifT^a_{adj}B^a_{\mu}g
\end{equation}
and $\bar{g}$, $g$ are the
$2(2S+1)$- dimensional WF of the color octet, $T^a_{adj}$ are the $SU(3)$
generators
in the adjoin representation. Employing the technique of  functional
integration and using a scheme which is analogous to the  $\bar{q}qg$- vertex
case, we suggest  that the interaction vertex of two structure gluons with the
gauge gluon has the following  analytical form\footnote{We did not allow for
the term $\delta_{\mu\nu}(p+k)_{\nu}$ which is the result of the auxiliary
condition (Klein-Gordon equation) to the Weinberg equation. In the
Kadyshevsky's approach all the particles, even in the intermediate states, are
on the mass shell.}:
\begin{equation}
i f\gamma^{\mu\nu}(p+k)_{\nu}(T^a_{adj})_{\alpha\beta}
\end{equation}
without  taking into account the multipole momenta  (compare with the QED
expression (\ref{22})).

Using results of  refs.~\cite{a14,a20}, let us  represent the Feynman matrix
element corresponding to the diagram of one-gluon exchange (see {\it Fig. 1})
as:
\begin{eqnarray}
\lefteqn{<p_1, p_2; \sigma_1, \sigma_2\vert \hat V^{(2)}\vert k_1, k_2;
\nu_1, \nu_2> = <p_1, p_2; \sigma_1, \sigma_2\vert \hat T^{(2)}\vert k_1. k_2;
\nu_1, \nu_2> =}\nonumber\\
&=&\sum^{1}_{\sigma_{ip}, \nu_{ip}, \nu_{ik} =-1}
D^{+\quad (S=1)}_{\sigma_1\sigma_{1p}} \left \{V^{- 1} (\Lambda_{\cal P},
p_1)\right \}
 D^{+\quad (S=1)}_{\sigma_2\sigma_{2p}}
\left \{V^{-1}(\Lambda_{\cal P}, p_2)\right \}\times\nonumber\\
&\times&V^{\nu_{1p}\nu_{2p}}_{\sigma
_{1p}\sigma_{ 2p}}(\vec k(-)\vec p, \vec p) D^{(S=1)}_{\nu_{1p}\nu_{1k}}\left
\{
V^{-1} (\Lambda_{p_1}, k_1)\right \} D^{(S=1)}_{\nu_{1k}\nu_1}\left \{
V^{-1}(\Lambda_{\cal P}, k_1)\right \}\times\nonumber\\
&\times& D^{(S=1)}_{\nu_{2p}\nu_{2k}} \left\{ V^{-1} (\Lambda_{p_2},
k_2)\right \} D^{(S=1)}_{\nu_{2k}\nu_2}\left\{ V^{-1} (\Lambda_{\cal P},
k_2)\right \},
\end{eqnarray}
where
\begin{equation}
V^{\nu_{1p}\nu_{2p}}_{\sigma_{1p}\sigma_{2p}} (\vec k(-) \vec p, \vec p) =
\xi_{\sigma_{1p}} \xi_{\sigma_{2p}} \hat V^{(2)} (\vec k(-) \vec p,
\vec p) \xi_{\nu_{1p}} \xi_{\nu_{2p}}.
\end{equation}
After  some  calculation, using the formulae of ref.~\cite{a14,Chesh}
\begin{equation}
U_\sigma(\vec p)=S_{\vec p} U_\sigma(\vec 0),\quad S_{\vec p}^{-1}S_{\vec
k}=S_{\vec k(-)\vec p}\cdot I\otimes D^{1}\left (V^{-1}(\Lambda_{\vec p}, \vec
k)\right ),
\end{equation}
\begin{equation}
S_{\vec p}^{-1}\gamma_{\mu\nu}p_\nu S_{\vec p}=\gamma_{44}\left (p_{\mu}
-\gamma_5 W_\mu (\vec p)\right ),
\end{equation}
\begin{equation}
W_\mu(\vec p)\cdot D\left (V^{-1}(\Lambda_{\vec p},\vec k)\right )= D\left (
V^{-1}(\Lambda_{\vec p},\vec k)\right )\cdot\left \{W_\mu(\vec
k)+\frac{p_\mu+k_\mu}{M(\Delta_0+M)}p_\nu W_\nu (\vec k)\right \},
\end{equation}
\begin{equation}
k_\mu W_\mu (\vec p)\cdot D\left (V^{-1}(\Lambda_{\vec p},\vec k)\right )=-
D\left (V^{-1}(\Lambda_{\vec p},\vec k)\right )\cdot p_\mu W_\mu (\vec k),
\end{equation}
we come to the 4- current of vector particle\footnote{$W_\mu$ is the
Pauli-Lyuban'sky 4- vector of relativistic spin.}:
\begin{equation}
j_\mu^{\sigma_{p}\nu_{p}}(\vec p, \vec k)= -f \xi_{\sigma_p} \left \{
(p+k)_{\mu}+{1\over M}W_{\mu}(\vec p)(\vec S\vec \Delta)-{1\over M}(\vec
S\vec\Delta)W_{\mu}(\vec p)\right \} \xi_{\nu_p}
\end{equation}
and  to the quasipotential
\begin{eqnarray}
\lefteqn{\hat V^{(2)} (\vec k(-)\vec p, \vec p) = -3f^2 \left\{ \frac{\left
[p_0
(\Delta_0 +M) + (\vec p
\vec\Delta)\right ]^2 -M^3 (\Delta_0+M)} {M^3 (\Delta_0 -M)}
+\right.}\nonumber\\
&+&\left.\frac{i (\vec S_1+\vec S_2)\left [\vec p\vec \Delta\right ]} {\Delta_0
-M}
\left [ \frac{p_0 (\Delta_0 +M)+\vec p\vec\Delta}{M^3}
\right ] + \frac{(\vec S_1\vec\Delta)(\vec S_2\vec\Delta)-(\vec
S_1\vec S_2)\vec\Delta^2}{2M(\Delta_0-M)}-\right.\nonumber\\
&-&\left.\frac{1}{M^3}\frac{\vec S_1\left [\vec p \vec \Delta\right ]\vec S_2
\left [\vec
p\vec\Delta\right ]}{\Delta_0-M}\right\}. \label{212}
\end{eqnarray}
As used in the earlier works~\cite{a11,a20}, we have
\begin{equation}
\vec\Delta = \Lambda^{-1}_{\vec p} {\vec k} = \vec k(-) \vec p = \vec
k-\frac{\vec p}{M} (k_0 - \frac{\vec k\vec p}{p_0 +M}),
\end{equation}
\begin{equation}
\Delta_0 = (\Lambda^{-1}_{\vec p} k)_0 = (k_0 p_0 -\vec k\vec p)/M,
\end{equation}
where $\vec{p}$, $\vec{k}$ are the covariant generalizations\footnote{We omit
the circles above
the covariant generalizations of the momenta,  as opposed
to~\cite{a14,a25,a20}.} of
the vectors of particle momenta in c.m.s.,  before $\vec{p}_1=-\vec{p}_2
=\vec{p}$ and after $\vec{k}_1=-\vec{k}_2=\vec{k}$ interaction;
$\xi^*$, $\xi$ are the analogues of  Pauli spinors  and $D^J_{\alpha\beta}$ is
the Wigner matrix of the irreducible representation of the rotation group,
which has  dimension equal to $(2S+1)$ with the following form:
\begin{eqnarray}
\lefteqn{D^{(S=1)}\left \{ V^{-1}(\Lambda_{\vec p}, \vec k)\right \}=
\frac{1}{2M(p_0+M)(k_0+M)(\Delta_0+M)} \left \{ \left [\vec p\vec k\right
]^2+\right.}\nonumber\\
&+&\left.\left [(p_0+M)(k_0+M)
-\vec k\vec p\right ]^2 -2i\left [(p_0+M)(k_0+M)-\vec k\vec p\right ] \vec
S\left [\vec p\vec
k\right ]-\right.\nonumber\\
&-&\left.2\{\vec S\left [\vec p\vec k\right
]\}^2\right \}.
\end{eqnarray}

The expression (\ref{212}) shows   the advantages of the $2(2S+1)$-
formalism, since it looks like  the quasipotential for the interaction of two
spinor particles with the substitutions
$\frac{1}{2m(\Delta-m)}\Rightarrow\frac{1}{\vec{\Delta}^2}$ and
$\vec{S}\Rightarrow\vec{\sigma}$.

\setcounter{equation}{0}
\section{System of two-particle partial-wave equations
in the relativistic configurational representation}

\hspace*{8mm}The transformations into the relativistic configurational
representation (RCR) have
the following form:
\begin{equation}
V(r,\vec n; \vec p) = {1\over (2\pi)^{3}} \int d\Omega_{\Delta}\, \xi^{*} (\vec
\Delta; \vec r) V (\vec \Delta, \vec p), \label{31}
\end{equation}
for the quasipotential and
\begin{equation}
\Psi_{\sigma_1\sigma_2} (r,\vec n) = {1\over (2\pi)^{3}} \int d\Omega_p\, \xi
(\vec p; \vec r) \Psi_{\sigma_1\sigma_2} (\vec p), \label{32}
\end{equation}
for the WF. The integration measure is
\begin{equation}
d\Omega_p \equiv d^3\vec p /\sqrt{1+\vec p^{\,2}/M^2}
\end{equation}
It is the invariant measure on the hyperboloid, $p_0^2-\vec p^{\,2}=M^2$. The
system of functions  $\xi(\vec{p};\vec{n},r)$ is the complete orthogonal system
of functions in the Lobachevsky space,
\begin{equation}
\xi (\vec p; \vec n; r) = (\frac{p_0-\vec p\vec n}{M})^{-1-irM}.
\end{equation}
The physical meaning of the parameter $r$ is discussed in details in
ref.~\cite{a21}.

As a result of  carring out  this transformation to the RCR we arrive at  the
following quasipotential:
\begin{equation}
V(r,\vec n; p_0, \vec p) = V_1 (r, p_0) + V_2 (r,\vec n; p_0, \vec
p),
\end{equation}
with
\begin{eqnarray}
\lefteqn{V_1 (r; p_0) = - 3f^2 \left \{ (8p^2_0 -4M^2) V_{Yuk} (r) +
(\frac{3p^2_0}{M^2}-1) \frac{1}{r} \delta
(r^2+\frac{1}{M^2})+\right.}\nonumber\\
&+&
\left.\frac{p_0^2}{M^2} \frac{1}{r}\delta (r^2+ \frac{4}{M^2}) +\frac{2 \vec
p^{\,2}}
{M^2} \left [B(r) + \frac{1}{3} \frac{1}{r}\delta (r^2 + \frac{1}{M^2})\right
]-\right.\nonumber\\
&-&\left.
(\vec S_1 \vec S_2) \left [\frac{2 \vec p^{\,2}}{M^2} B(r) + \frac {1}{6} \frac
{1}{r}
\delta (r^2 + \frac{1}{M^2}) +\frac{1}{2}\frac{1}{r} \delta (r^2+
\frac{4}{M^2})\right ]\right \},
\end{eqnarray}
and
\begin{eqnarray}
\lefteqn{V_2(r, \vec n; p_0, \vec p)=-3f^2 \left \{ \frac{2}{M^2}(\vec S_1 \vec
p)
(\vec S_2 \vec p) B(r) - S_{12} B(r) + \right.}\nonumber\\
&+&\left. \frac{3}{M^2} \left [ (\vec S_1 \vec L)(\vec S_2 \vec L) + (\vec S_2
\vec
L)(\vec S_1\vec L)\right ] \frac{1}{r^2} B(r) - \frac {6}{M^2}(\vec p
\vec n)^2 B(r) + \right.\nonumber\\
&+&\left. \frac{2ip_0}{M} (\vec p \vec n) \left [4rA(r) - \frac {1}{M^2}
C(r)\right ] -
(\vec S \vec L) \left [ \frac {4 p_0}{M} A(r) +\right.\right.\nonumber\\
&+&\left.\left.
\frac{6i}{M^2} (\vec p \vec n) \frac {1}{r} B(r) - \frac {p_0}{M^3}
\frac {1}{r} C(r)\right ]\right \},
\end{eqnarray}
where $\vec{S}=\vec{S}_1+\vec{S}_2$, $\vec{L}=[\vec{p}\times\vec{r}]$,
$S_{12}=3(\vec{S}_1\vec{n})(\vec{S}_2\vec{n})-(\vec{S}_1\vec{S}_2)$.

Next,
\begin{equation}
V_{Yuk} (r)= \frac{1}{4\pi r} cth (r M \pi),
\end{equation}
and
\begin{eqnarray}
A(r)&=&\frac{1}{r(r+(i/M))} V_{Yuk}(r), \\
B(r)&=& \frac{1}{(r+(i/M))(r+(2i/M))} V_{Yuk}(r), \\
C(r)&=&\frac{1}{(i/M)} \frac{(r - (i/M))}{r+(i/M)} \frac {1}{r} \delta
(r^2 + \frac {4}{M^2}) - \frac {M^2}{4r} \delta(r).
\end{eqnarray}

{}From the above equations we can see that the quasipotential is separated into
two parts, $V_1(r; p_0)$, which does not depend on the direction of the
''relativistic coordinate'' vector, and $V_2(r, \vec{n}; p_0, \vec{p})$, which
depends on $\vec{n}$ over the structure
$(\vec{p}\vec{n})$, $\left [\vec{p}\times\vec{n}\right ]$, $S_{12}$.

After the transformation of the quasipotential equations:
\begin{equation}
({\cal M} - 2p_0) \Psi_{\sigma_1 \sigma_2}(\vec p)=(2\pi)^{-3}
\sum_{\nu_1\nu_2}
\int d\Omega_k V^{\nu_1\nu_2}_{\sigma_1\sigma_2} (\vec k, \vec p)
\Psi_{\nu_1\nu_2} (\vec k), \label{312}
\end{equation}
in the RCR by means of the formulae (\ref{31},\ref{32}),
we can be convinced that $V_1(r; p_0)$ describes the local type of interaction
and $V_2(r, \vec{n}; p_0, \vec{p})$ enters into equation (\ref{eq:rcr27}) by
the non-local way.
\begin{eqnarray}\label{eq:rcr27}
\lefteqn{({\cal M} -2\hat H)\Psi_{\sigma_1\sigma_2}(\vec r)=\sum_{\nu_1\nu_2}
V_{1_{\sigma_1\sigma_2}}^{\nu_1\nu_2} (r; p_0)\Psi_{\nu_1\nu_2}(\vec r)
+ \int d^3 \vec r_1\sum_{\nu_1\nu_2} \int d\Omega_p \times}\nonumber\\
&\times& \xi(\vec p; \vec n, r)\xi(\vec p; \vec n_1, r_1)
V^{\nu_1\nu_2}_{2_{\sigma_1\sigma_2}} (r_1,\vec n_{1\Lambda p}; p_0,\vec
p)\Psi_{\nu_1\nu_2}(\vec r_1),\label{313}
\end{eqnarray}
where the unit vector is~\cite{a22}
\begin{equation}
\vec n_{\Lambda p}=[M\vec n - \vec p (1 - \frac{\vec p \vec n}{p_0+M})]/
(p - \vec p \vec n).
\end{equation}
It is still possible to localize the spin-orbit part, some terms of the tensor
interaction and some other terms entering in $V_2(r_1, \vec{n}_{1\Lambda_p};
p_0, \vec{p})$.

We have the following structure in  Eq. (\ref{313}):
\begin{eqnarray}
(\vec p \cdot\vec n_{1\Lambda p})&=& M^2/(p_0 - \vec p\vec n_1) - p_0,
\label{315}\\
\left [\vec p \times \vec n_{1\Lambda p}\right ]&=& M\left [\vec p \times \vec
n_1\right ]/(p_0-\vec p\vec n_1), \label{316}
\end{eqnarray}
and
\begin{equation}
(\vec S_1 \vec n_{1\Lambda p}) (\vec S_2 \vec n_{1\Lambda p}) = Z^T_1+ Z^T_2,
\end{equation}
where
\begin{eqnarray}
Z^T_1&=&M^2(\vec S_1 \vec n_1)(\vec S_2 \vec n_1)/(p_0 - \vec p \vec
n_1), \\
Z^T_2&=&\frac{M^2}{(p_0 -\vec p \vec n_1)^2} \{- \frac {1}{M}[(\vec S_1 \vec
n_1)
(\vec S_2 \vec p) + (\vec S_1 \vec p)(\vec S_2 \vec n_1)] \times\nonumber\\
&\times& (1 - \frac {\vec p \vec n_1}{p_0 + M}) +
\frac {1}{M^2}(\vec S_1 \vec p)
(\vec S_2 \vec p)(1 - \frac {\vec p \vec n_1}{p_0 + M})^2\}.
\end{eqnarray}

The localization procedure  for (\ref{315},\ref{316}) and the first term of
$Z^{T}_{1}$ is produced by means of the following equation:
\begin{equation}
\exp(\frac{i}{M} \frac {\partial}{\partial r_1}) \xi^{*}(\vec p; \vec
n_1,r_1) =\frac {M}{p_0 - \vec p \vec n_1} \xi^{*} (\vec p; \vec
n_1,r_1).
\end{equation}
As a result we obtain
\begin{equation}
({\cal M} - 2\hat H)\Psi_{\sigma_1\sigma_2}(\vec r)=\sum_{\nu_1\nu_2}
\hat V^{\nu_1\nu_2}_{\sigma_1\sigma_2}(\vec r; p_0,\vec p)\Psi_{\nu_1\nu_2}
(\vec r).
\end{equation}

The quasipotential of  Eq. (\ref{eq:rcr27}) is presented as a sum of six
component:
\begin{equation}
\hat V(\vec r; p_0,\vec p)=\hat V_C +\hat V_{LS} (\vec L \vec S)+
\hat V_S( \vec S_1 \vec S_2) + \hat V_TS_{12}+\hat V_{LL}L_{12}+\hat V_{Sp}
P_{12}, \label{322}
\end{equation}
where
\begin{eqnarray}
L_{12}&=&\frac {1}{2} \{(\vec S_1 \vec L)(\vec S_2 \vec L)+(\vec S_2 \vec L)
(\vec S_1 \vec L)\}, \\
P_{12}&=&(\vec S_1 \vec p)(\vec S_2 \vec p);
\end{eqnarray}
and
\begin{eqnarray}
\hat V_C&=&-3f^2\left \{ (8p_0-4M^2) V_{Yuk}(r)+(\frac{3p_0^2}{M^2} -1)
\frac {1}{r}\delta (r^2+ \frac {1}{M^2})+ \frac {p_0^2}{M^2} \frac{1}{r}\delta
(r^2 + \frac {4}{M^2})-\right.\nonumber\\
&+&\left. 2\frac{\vec p^2}{M^2}\left [B(r) +\frac{1}{3} \frac {1}{r} \delta
(r^2+ \frac{1}{M^2})\right ] - 6\left [\frac{(r- \frac{2i}{M})^2}{r^2}
\exp(-\frac{2i}{M}
\frac{\partial}{\partial r})- \frac {2p_0}{M} \frac {(r- \frac
{i}{M})^2}{r^2} \times\right.\right.\nonumber\\
&\times&\left.\left.\exp(- \frac{i}{M} \frac{\partial}{\partial r})+
\frac{p_0^2}{M^2}\right ]B(r)+ \frac {2ip_0}{M}\left [M \frac{(r - \frac
{i}{M})^2}{r^2}
\exp(- \frac{i}{M} \frac{\partial}{\partial r}) -p_0\right
]\right.\times\nonumber\\
&\times&\left.\left [4rA(r)- \frac{1}{M^3} C(r)\right ]\right \},\label{325}\\
\hat V_{LS}&=& - 3f^2\left \{ \frac{p_0}{M} \frac{r^2}{(r+(i/M))^2} \left
[4A(r)
- \frac{1}{M^2} \frac {1}{r} C(r)\right ]+\right.\nonumber\\
&+&\left.\frac {6i}{M^2} \left [M \frac {r(r-(i/M))}{(r+(i/M))^2}   \exp (-
\frac{i}{M} \frac {\partial}{\partial r}) - p_0\right ] \frac{r}{(r+(i/M))^2}
B(r)\right \}, \\
\hat V_S&=&-3f^2\left \{-\left [\frac {(r-(2i/M))^2}{r^2}  \exp (- \frac{2i}{M}
\frac
{\partial} {\partial r}) + \frac {2\vec p^2}{M^2} - 1\right ] B(r)
-\right.\nonumber\\
&-&\left. \frac {1}{6} \frac{1}{r} \delta (r^2+ \frac {1}{M^2}) - \frac{1}{2}
\frac {1}{r} \delta (r^2 + \frac {4}{M^2})\right \}, \\
\hat V_T&=&-3f^2\left \{ - \frac {(r -(2i/M))^2}{r^2} \exp (- \frac {2i}{M}
\frac
{\partial}{\partial r}) B(r)\right \}, \\
\hat V_{LL}&=& -3f^2 \left \{ \frac{3}{M^2} \frac{r}{(r+(i/M))(r+(2i/M))^2}
B(r)\right \}, \\
\hat V_{Sp}&=&-3f^2 \{ \frac{2}{M^2} B(r)\}. \label{330}
\end{eqnarray}

It is noted that we have assumed the corresponding finite-difference operators
in $p_0$ and $\vec{p}$  for these calculations. However, the use of eigenvalues
instead of these operators is a  good approximation, which does not reduce to
the non-relativistic models.

After carrying out the partial-wave expansion for the WF
\begin{equation}
\Psi^{(S)}(\vec r; \sigma)=\frac{4\pi}{r}
\sum_{J\ell M}
R_{J\ell S} (r) \left \{ \Omega^{\ast (S)}_{J \ell
M}(\vec n) \right \}_{\sigma},
\end{equation}
the three-dimensional quasipotential equation is rewritten in the system of
radial equations for $S=0$, $S=1$ and $S=2$
\begin{equation}
({\cal M}-2 \hat H_{\ell})R_{J\ell S}(r)=
\sum_ {\ell^{\prime} S^{\prime}}\hat V^J_{\ell S,
\ell^{\prime} S^{\prime}} (r; p_0,\vec p)
R_{J\ell^{\prime} S^{\prime}}(r). \label{332}
\end{equation}
Here,
\begin{equation}
\hat H_\ell=M\,cosh(\frac{i}{M} \frac{\partial}{\partial r}) + \frac
{i}{r} sinh (\frac {i}{M} \frac{\partial}{\partial r}) + \frac
{\ell(\ell+1)}{2Mr^2} \exp (\frac{i}{M} \frac{\partial}{\partial r}),
\end{equation}
and
\begin{equation}
\hat V^J_{\ell^\prime S^\prime, \ell S} (r; p_0,\vec p)= \int
d\omega_{\vec n} \Omega^{\ast (S^\prime)}_{J \ell^\prime M} (\vec n) \hat
V (\vec r; p_0 , \vec p) \Omega^{(S)}_{J \ell M} (\vec n).\label{334}
\end{equation}

One can check that  due to the complete accordance of the relativistic spin
structure of the quasipotential (\ref{322}) to the spin structures
which are used in non-relativistic models, the matrix elements (\ref{334}) have
the same form as when non-relativistic interaction of two vector particles is
considered. The only differences are in
the internal expressions of $\hat{V}_C$, $\hat{V}_T$, $\hat{V}_{SL}$,
$\hat{V}_S$, $\hat{V}_{LL}$ and $\hat{V}_{Sp}$.

Thus, we have for a singlet state
\begin{equation}
({\cal M}-2 \hat H_{\ell=J})R_{\ell=J} (r)=\left (\hat V_C
-2\hat V_S - \frac
{2}{3}J (J+1) \hat V_{LL}\right ) R_{\ell=J}(r); \label{335}
\end{equation}
for a triplet state
\begin{eqnarray}
&&({\cal M}-2 \hat H_{\ell=J+1})R_{\ell=J+1} (r)=\left (\hat
V_C
-(J+2)\hat V_{SL}  - \hat V_{S} - \frac {J+2}{2J+1} \hat V_T + \right.
\nonumber\\
&&\left.\qquad + \frac
{J+2}{2} \hat V_{LL}\right ) R_{\ell=J+1}(r)+ \frac {3\sqrt{
J(J+1)}}{2J+1} \hat V_T R_{\ell = J-1} (r), \\
&&({\cal M}-2\hat H_{\ell=J})\quad R_{\ell =J} (r) = \left (\hat
V_C-\hat
V_{LS}-\hat V_S + \hat V_T+ [\frac{1}{2}-J(J+1)]\hat V_{LL}\right )
R_{\ell=J} (r), \nonumber\\
&&\\
&&({\cal M}-2\hat H_{\ell=J-1}) R_{\ell=J-1} (r)=\left (\hat V_C
+(J-1)\hat V_{LS}  - \hat V_{S} - \frac {J-1}{2J+1} \hat V_T - \right.
\nonumber\\
&&\left.\qquad - \frac{J-1}{2} \hat V_{LL}\right  ) R_{\ell=J-1} (r) +
 \frac{3\sqrt{ J(J+1)}}{2J+1} \hat V_T
R_{\ell = J+1} (r);
\end{eqnarray}
for 5-plet state
\begin{eqnarray}
&&({\cal M}-2\hat H_{\ell=J+2} )R_{\ell=J+2} (r)=\left (\hat V_C
-2(J+3)\hat V_{LS} + \hat V_{S} - \frac {2(J+3)}{2J+3} \hat V_T + \right.
\nonumber\\
&&\left.\qquad +(J+3)^2 \hat V_{LL}\right ) R_{\ell=J+2}(r)+
\frac{\sqrt{ 6J(J+2)}}{2J+1}
\frac {\sqrt{ (2J-1)(2J+1)}}{2J+3} \hat V_T R_{\ell = J}(r),
\nonumber\\
&&\\
&& ({\cal M}-2\hat H_{\ell=J+1} ) R_{\ell=J+1} (r)=\left (\hat
V_C
-(J+4)\hat V_{LS}  + \hat V_{S} +\frac {J-4}{2J+1} \hat V_T +  \right.
\nonumber\\
&&\left.\qquad + \frac {3J+8}{2} \hat V_{LL}\right ) R_{\ell=J+1}(r)+
\frac {3\sqrt{ (J-1)(J+2)}}{2J+1} \hat V_T
R_{\ell = J-1} (r), \\
&& ({\cal M}-2\hat H_{\ell=J} ) R_{\ell=J} (r)=
\frac{\sqrt{6J(J+2)}}{2J+1} \frac{\sqrt{(2J-1)(2J+1)}}{2J+3}
\hat V_T R_{\ell=J+2}(r)+ \nonumber\\
&&\qquad +\left  (\hat V_C- 3\hat V_{LS} +\hat V_{S}+\frac
{(2J-3)(2J+5)}{(2J-1)(2J+3)} \hat V_T +\left [\frac{5}{2}-\frac{1}{3}J
(J+1)\right ]\hat V_{LL}\right ) \times
\nonumber\\
&&\qquad \times R_{\ell=J}(r) +\frac{\sqrt{ 6(J-1)(J+1)}}{2J+1}
 \frac{\sqrt{(2J+1)(2J+3)}}{2J-1} \hat V_T R_{\ell = J-2} (r),\\
&&({\cal M}-2\hat H_{\ell=J-1}) R_{\ell=J-1} (r)=\left (\hat V_C
+(J-3)\hat V_{LS}  + \hat V_{S} + \frac {J+5}{2J+1} \hat V_T -
\right.\nonumber\\
&&\left.\qquad - \frac{3J-5}{2} \hat V_{LL}\right ) R_{\ell=J-1}(r)+
\frac {3\sqrt{ (J-1)(J+2)}}{2J+1} \hat V_T R_{\ell = J+1} (r), \\
&&({\cal M}-2\hat H_{\ell=J-2}) R_{\ell=J-2} (r)=\left (\hat V_C
+2(J-2)\hat V_{LS}  + \hat V_{S} - \frac {2(J-2)}{2J-1} \hat V_T + \right.
\nonumber\\
&&\left. \qquad+ (J-2)^2  \hat V_{LL}\right ) R_{\ell=J-2}(r)+
\frac{\sqrt{6(J-1)(J+1)}}{2J+1} \frac{\sqrt{(2J+1)(2J+3)}}{2J-1}
\hat V_{T}R_{\ell=J} (r).\nonumber\\
&& \label{343}
\end{eqnarray}

The two-gluon bound states are to have the positive $C$- parity. Therefore, we
have  restrictions on $J$ in (\ref{335}-\ref{343}).

Finally, let us remark that we have neglected the last term of (\ref{322})
following  the authors of ref.~\cite{a23}.

\setcounter{equation}{0}
\section{Quasiclassical condition for a quantization in  RCR}

\hspace*{8mm}The non-tied partial-wave equations can be re-written in the
following form:
\begin{equation}
[cosh (i\bar\lambda \frac{\partial}{\partial r}) + \frac{i\bar\lambda}{r} sinh
(i\bar\lambda \frac{\partial}{\partial r})
+ \frac{\bar\lambda^2\ell (\ell
+1)}{2r^2} \exp (i\bar\lambda \frac{\partial}{\partial r}) - X (r)]
R_\ell (r) = 0,
\end{equation}
\begin{equation}
X(r) = \frac{W-V(r)}{2M},
\end{equation}
where $W=2M+E_{b} $ is the mass of bound state, $M$ is the gluon mass, $E_{b}$
is the
binding energy. We consider $V(r)$ as a sum of the potential $V_{conf}$
describing the confinement of a gluon in a meson and the corresponding matrix
elements of the quasipotential in the system of the Eqs. (\ref{332}).
In the expressions (\ref{325}-\ref{330})
it is possible to neglect the small image additions which are proportional to
$i/M(=i\hbar/Mc)$, because the bound state spectrum usually forms on  distances
of the order $r>>\bar\lambda=\hbar/Mc$.

The quasiclassical condition for a quantization of the two-particle
relativistic systems has the following form~\cite{a12}:
\begin{equation}
\int^{r_+}_{r_-} dr^\prime Arch X_\Lambda (r^\prime) = \bar\lambda \pi
(n +\frac{1}{2} ),
\end{equation}
where
\begin{equation}
X_\Lambda (r) = X(r) [1+(\Lambda \frac{\bar\lambda}{r})^2]^{-1/2},\quad \Lambda
=\ell+1,
\end{equation}
and the integration limits are determined from the equation
\begin{equation}
X_{\Lambda} (r_\pm) = 1.
\end{equation}

If  restrict ourselves to the cases of the simplest potentials, namely,
$V(r)=\sigma r$ and $V(r)=\sigma r^2$, the quantization condition is shown in
ref.~\cite{a24} to take the
 form of (\ref{46}) and (\ref{47}), correspondingly,
\begin{equation}
\chi\,\, cosh\,\chi - sinh\,\chi = \frac{\sigma}{2Mc^2} \bar\lambda \pi
(n+\frac{\ell}{2}+\frac{3}{4}),   \label{46}
\end{equation}
\begin{equation}
2\sqrt{cosh\,\chi +1} \left [K(tanh\, \chi/2) - E(tanh\, \chi/2) \right ] =
\sqrt{
\frac{\sigma}{2Mc^2}} \bar\lambda \pi (n+\frac{\ell}{2}+\frac{3}{4}),
\label{47}
\end{equation}
where $cosh\, \chi=W/2M$.
Following  this technique, the energy levels of gluonium states can be
obtained.
The numerical results for the case of the confinement potential are presented
in ref.~\cite{a25}.
Investigation of the influence of the Coulomb term as well as the relativistic
corrections are
in progress.\\

\section{Conclusions}

\hspace*{8mm}In the present work the formalism for consideration of the two
vector particles, gluons, has been constructed on the basis of the $2(2S+1)$-
dimensional description of the WF. The form of the relativistic two-particle
single-time quasipotential equation was found. It  turned out that the
quasipotential of this equation coincided with the quasipotential for
interaction of two spinor particles in the second order of perturbation theory.
This fact shows the advantages  of the $2(2S+1)$- component formalism.

It is possible to employ  the relativistic generalization of the WKB method  to
find the gluonium spectrum in the case when the total spin of the system is
equal to zero.
In  following works we will use the obtained system of the partial-wave
equations for
investigation of  the contributions of the spin-spin and
spin-orbit interactions to the energy of the gluonium states.

{\bf {Acknowledgements.}} We would like to express my sincere gratitude to
Profs.
A. M. Cetto, M. Moreno,  N.~B.~ Skachkov, M. Torres, Yu. N.Tyukhtyaev and
C. Villareal for interest in the work and for very helpful discussions.  The
assistance of  Drs.
K. Michaelian and  R. Garc\'{\i}a-Pelayo   is greatly appreciated.

This work has been financially supported by the CONACYT (Mexico) under contract
No. 920193.

\end{document}